\documentclass[12pt,a4paper]{article}

\usepackage[utf8]{inputenc}
\usepackage[T1]{fontenc}
\usepackage{lmodern}
\usepackage[margin=2.5cm]{geometry}
\usepackage{microtype}

\usepackage{etoolbox}
\makeatletter
\patchcmd{\showhyphens}{\reset@font}{}{}{}
\makeatother

\usepackage{amsmath,amssymb,amsfonts}
\usepackage{amsthm}

\usepackage{graphicx}
\usepackage{multirow}
\usepackage{booktabs}
\usepackage{xcolor}

\usepackage{algorithm}
\usepackage{algorithmicx}
\usepackage{algpseudocode}
\usepackage{listings}
\usepackage{textcomp}

\usepackage[title]{appendix}
\usepackage{hyperref}
\usepackage{orcidlink}   

\hypersetup{
    colorlinks=true,
    linkcolor=blue!60!black,
    citecolor=blue!60!black,
    urlcolor=blue!60!black,
    pdftitle={Automatic Identification of Parallelizable Loops Using Transformer-Based Source Code Representations},
    pdfauthor={Izavan dos S. Correia, Henrique C. T. Santos, Tiago A. E. Ferreira}
}

\theoremstyle{plain}

\theoremstyle{definition}

\theoremstyle{remark}

\usepackage[numbers,sort&compress]{natbib}

\usepackage{setspace}
\usepackage{authblk}

\begin{document}

\title{\textbf{Automatic Identification of Parallelizable Loops Using\\
Transformer-Based Source Code Representations}}

\author[1]{Izavan dos S. Correia\,\orcidlink{0000-0002-5757-4473}}
\author[2]{Henrique C. T. Santos\,\orcidlink{0000-0002-9544-7774}}
\author[1]{Tiago A. E. Ferreira\,\orcidlink{0000-0002-2131-9825}}

\affil[1]{Graduate Program in Applied Informatics (PPGIA),
Federal Rural University of Pernambuco (UFRPE), Recife, Pernambuco, Brazil.
\texttt{izavan.correia@ufrpe.br}, \texttt{tiago.espinola@ufrpe.br}}

\affil[2]{Undergraduate Program in Analysis and Systems Development (TADS),
Federal Institute of Pernambuco (IFPE), Recife, Pernambuco, Brazil.
\texttt{henrique.santos@recife.ifpe.edu.br}}

\date{}  

\maketitle

\begin{abstract}
Automatic parallelization remains a challenging problem in software engineering, particularly in identifying code regions where loops can be safely executed in parallel on modern multi-core architectures. Traditional static analysis techniques, such as dependence analysis and polyhedral models, often struggle with irregular or dynamically structured code. In this work, we propose a Transformer-based approach to classify the parallelization potential of source code, focusing on distinguishing independent (parallelizable) loops from undefined ones. We adopt DistilBERT to process source code sequences using subword tokenization, enabling the model to capture contextual syntactic and semantic patterns without handcrafted features. The approach is evaluated on a balanced dataset combining synthetically generated loops and manually annotated real-world code, using 10-fold cross-validation and multiple performance metrics. Results show consistently high performance, with mean accuracy above 99\% and low false positive rates, demonstrating robustness and reliability. Compared to prior token-based methods, the proposed approach simplifies preprocessing while improving generalization and maintaining computational efficiency. These findings highlight the potential of lightweight Transformer models for practical identification of parallelization opportunities at the loop level.
\end{abstract}


\vspace{1em}
\noindent\textbf{Funding:} This work was supported by the Coordination for the
Improvement of Higher Education Personnel -- Brazil (CAPES) -- Finance Code 001
through a master's scholarship.

\vspace{1em}
\noindent\textbf{Conflict of interest:} The authors declare no conflicts of interest.

\newpage

\section{Introduction}
\label{sec:introduction}

Automatic parallelization remains a key challenge in software engineering and
compiler optimization, as it directly impacts application performance on modern
multi-core and heterogeneous architectures. Traditional static analysis
techniques, such as dependence analysis and polyhedral frameworks, rely on
precise code representations including abstract syntax trees (ASTs) and
control-flow graphs to determine whether loop iterations can be executed in
parallel without conflicts \citep{zinenko2018, pouchet2009}. However, these
approaches often face limitations when dealing with irregular, non-affine, or
highly dynamic code structures, which are common in real-world software systems.

Machine learning and deep learning have emerged as promising alternatives for
source code analysis by learning latent representations directly from raw code,
reducing the need for handcrafted rules and domain-specific
heuristics \citep{allamanis2018survey}. Early approaches primarily relied on
token-level or sequence-based representations combined with shallow classifiers,
while more recent methods leverage deep neural architectures capable of capturing
richer syntactic and semantic properties of code. These techniques have shown
effectiveness across a wide range of software engineering tasks, including code
summarization, defect prediction, clone detection, and program
classification \citep{allamanis2016convolutional, white2016deep}.

More recently, transformer-based models have significantly advanced
representation learning for both natural language and source code through
self-attention mechanisms that enable the modeling of long-range dependencies
within sequences \citep{vaswani2017attention}. Pre-trained transformer models
such as CodeBERT and GraphCodeBERT adapt this architecture to programming
languages by jointly modeling natural language and code tokens, as well as
incorporating data-flow information, achieving state-of-the-art performance in
tasks such as code search, clone detection, and program
classification \citep{feng2020codebert, guo2020graphcodebert}. Furthermore,
transformer architectures have also been explored in contexts directly related to
parallelization, demonstrating their potential to classify code regions suitable
for OpenMP-based parallel execution \citep{chen2024pragformer}.

Despite these advances, many existing approaches for detecting parallelizable
code still rely on structured representations, such as ASTs or graph-based
embeddings, which require complex preprocessing pipelines and substantial domain
expertise. Previous work from this research group proposed a lightweight approach
based on deep neural networks (DNNs) and convolutional neural networks (CNNs),
trained on token sequences generated by a Python preprocessing script, to
distinguish independent (parallelizable) loops from dependent ones in synthetic
datasets \citep{correia2025discovering}. While the results demonstrated that
token-level representations contain relevant information for parallelism
classification, the approach required external preprocessing steps and
dimensionality reduction techniques such as principal component analysis (PCA),
and did not exploit contextual representations of code.

In this work, we extend that foundation by leveraging transformer-based
contextual representations to improve the automatic classification of
parallelization potential in source code. Specifically, we adopt DistilBERT, a
distilled and computationally efficient variant of BERT that preserves much of
the representational power of larger transformer models while reducing model size
and inference cost \citep{sanh2019distilbert}. DistilBERT produces contextualized
embeddings that capture both syntactic patterns and semantic relationships
directly from raw code sequences, enabling effective learning without extensive
manual feature engineering and making it suitable for scenarios with constrained
computational resources when compared to larger code-specific models.

To assess the effectiveness of the proposed approach, we constructed a balanced
dataset combining synthetically generated loops and manually annotated real-world
code samples, and evaluated the model using rigorous 10-fold cross-validation. In
addition to overall accuracy, we emphasize metrics such as the false positive rate
(FPR), which is particularly relevant in parallelization scenarios where
incorrectly classifying dependent code as parallelizable may lead to unsafe or
inefficient execution. The experimental results indicate that the
transformer-based model achieves stable and competitive performance across
multiple evaluation metrics, while simplifying the preprocessing pipeline and
improving generalization when compared to the previous token-based approach.

The remainder of this paper is organized as follows:
Section~\ref{sec:related_works} reviews related work on parallelization and deep
learning for code analysis, Section~\ref{sec:methodology} describes the proposed
method, Section~\ref{sec:results} presents the experimental evaluation,
Section~\ref{sec:discussion} discusses the findings, and
Section~\ref{sec:conclusion} concludes the paper and outlines directions for
future work.

\section{Related Works}
\label{sec:related_works}

Research on automatic parallelization has traditionally been dominated by
compiler-based static analysis techniques. Classical approaches rely on
dependence analysis, control-flow analysis, and polyhedral models to determine
whether loop iterations can be safely executed in
parallel \citep{pouchet2009, zinenko2018}. While these methods provide strong
theoretical guarantees, they require affine loop bounds and regular memory access
patterns, which significantly limits their applicability to real-world programs
that exhibit dynamic control flow, pointer aliasing, and irregular data access
patterns.

To overcome these limitations, machine learning techniques have been increasingly
explored for source code analysis. Early work focused on learning from
token-level representations or handcrafted features extracted from code,
demonstrating that statistical models could capture useful properties related to
program behavior \citep{allamanis2016convolutional}. More comprehensive surveys
highlight the growing adoption of machine learning and deep learning methods
across software engineering tasks, including bug detection, clone detection, and
program classification, emphasizing their ability to generalize beyond rigid
syntactic rules \citep{allamanis2018survey, white2016deep}.

The introduction of transformer architectures marked a significant advancement in
code representation learning. By leveraging self-attention mechanisms,
transformers enable the modeling of long-range dependencies within sequences,
which is particularly relevant for source code
understanding \citep{vaswani2017attention}. Building on this paradigm, models
such as CodeBERT and GraphCodeBERT incorporate pre-training on large-scale code
corpora and, in some cases, explicit data-flow information, achieving
state-of-the-art performance on several downstream tasks, including code search,
clone detection, and program
classification \citep{feng2020codebert, guo2020graphcodebert}. Despite their
effectiveness, these models are typically large and computationally demanding,
which may limit their applicability in scenarios with constrained resources.

More closely related to this work are studies that apply deep learning techniques
to identify parallelizable regions in code. Recent transformer-based approaches,
such as PragFormer, have demonstrated the feasibility of using attention-based
models to classify code segments according to their suitability for OpenMP
parallelization \citep{chen2024pragformer}. These methods highlight the potential
of learned representations to support parallelization decisions without relying
solely on traditional compiler analyses.

Prior work from this research group proposed a lightweight deep learning approach
based on deep neural networks and convolutional neural networks trained on token
sequences generated via Python-based preprocessing
scripts \citep{correia2025discovering}. That study demonstrated that even shallow
token representations contain sufficient information to distinguish independent
from dependent loops. However, the approach required external preprocessing steps
and dimensionality reduction techniques, such as principal component analysis, and
did not exploit contextual representations capable of modeling long-range
dependencies within code.

In contrast to existing methods, the present work leverages a distilled
transformer model to balance representational power and computational efficiency.
By adopting DistilBERT, the proposed approach captures contextual syntactic and
semantic information directly from source code while maintaining a lightweight
architecture suitable for practical deployment. This positions the method as a
complementary alternative to both traditional static analysis techniques and
larger transformer-based models, addressing the need for efficient and accurate
identification of parallelization opportunities in modern software systems.

\section{Methodology}
\label{sec:methodology}

\begin{figure}[htbp]
    \centering
    \includegraphics[width=0.9\linewidth]{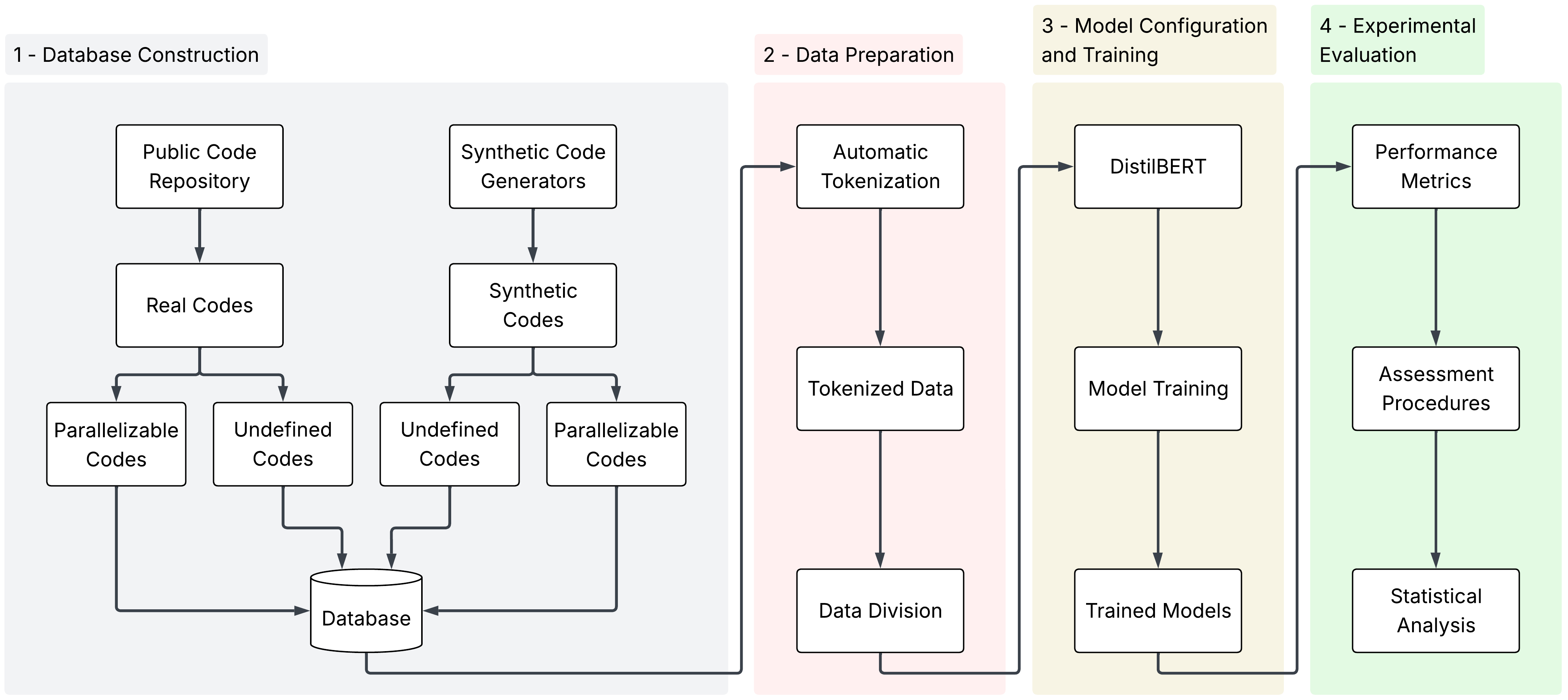}
    \caption{Workflow of the proposed methodology.}
    \label{fig:pipeline}
\end{figure}

This section describes the methodology for developing and evaluating the
automatic classification system for parallelization potential in programming
codes. The process involves five main stages: (1) database construction,
(2) data preparation, (3) model setup and training, (4) experimental evaluation,
and (5) computational environment. Figure~\ref{fig:pipeline} shows the complete
pipeline and the flow of each stage.

The process starts with the construction of the dataset using real and synthetic
code samples. Next, the data are automatically tokenized and split into training,
validation, and test sets. The \textit{DistilBERT} model is then configured and
fine-tuned using the prepared data. Finally, the trained models are evaluated
using performance metrics and statistical analyses.

\subsection{Database Construction}
\label{subsec:database}

The database was carefully built from two complementary sources: automatically
generated synthetic codes and real-world codes collected from public repositories.

\subsubsection{Synthetic Code Generation}
\label{subsubsec:synthetic_codes}

To ensure diversity and control over the classes of interest, we used two
generators based on genetic algorithms, following the methodology established in
our previous work \citep{correia2025discovering}. The generators were implemented
in Python using the DEAP (Distributed Evolutionary Algorithms in Python)
library \citep{deap2012} and produced two distinct types of codes:

\begin{itemize}
    \item \textbf{Class 1 -- Parallelizable}: Independent loops where each
    iteration has no data dependencies with other iterations, allowing parallel
    execution without conflicts.
    \item \textbf{Class 0 -- Undefined}: Loops with non-trivial dependencies or
    whose parallelization analysis requires complex considerations, including cases
    where parallelization is not possible or cannot be straightforwardly
    determined.
\end{itemize}

The evolutionary process employed a population of 10,000 individuals over 50
generations, with a crossover rate of 0.9 and a mutation rate of 0.1. The fitness
function considered multiple structural aspects of the code, including the number
of functions, conditionals, variables, loops, and the guarantee of successful
compilation. A total of 8,000 synthetic samples were generated, with the
distribution between classes determined by the evolutionary process.

\subsubsection{Real Code Collection and Annotation}
\label{subsubsec:real_codes}

To complement the synthetic data and enhance the database's representativeness,
we collected 340 real code samples from public GitHub repositories and examples
available on the internet. Each real code sample underwent a thorough manual
analysis conducted by the authors and was classified using the same criteria
defined in Section~\ref{subsubsec:synthetic_codes}:

\begin{itemize}
  \item \textbf{Class 1 -- Parallelizable}
  \item \textbf{Class 0 -- Undefined}
\end{itemize}

This manual analysis ensured the quality of the annotations and the relevance of
the examples for the classification task.

\subsubsection{Final Database Composition}
\label{subsubsec:database_composition}

The final database totaled 8,340 examples, consisting of:

\begin{itemize}
    \item \textbf{4,140 parallelizable codes}: comprising 4,000 synthetic samples
    and 170 real samples.
    \item \textbf{4,140 undefined codes}: comprising 4,000 synthetic samples and
    170 real samples.
\end{itemize}

After balancing the dataset, both classes (Parallelizable and Undefined) are
equally represented, ensuring a more uniform learning process and reducing
potential bias during training. No resampling techniques were required in the
final configuration, as the balanced distribution itself provided a robust
foundation for model training. With this updated dataset, the models achieved
excellent and stable performance, indicating that the class balance contributed
positively to the predictive reliability.

This research assumes that the synthetically generated loops cover representative
patterns of the repetition structures observed in real codes. More complex loop
transformations---such as fusion, unrolling, or tiling---were not included, as
these variations fall outside the scope of this work.

\subsection{Data Preparation}
\label{subsec:data_preparation}

Unlike the previous work \citep{correia2025discovering}, which relied on an
external Python script for automated tokenization and required dimensionality
reduction via Principal Component Analysis (PCA) \citep{pearson1901liii,
Gray2017}, the selected Transformer model processes source code directly as text,
significantly simplifying the preparation pipeline.

\subsubsection{Automatic Tokenization}
\label{subsubsec:tokenization}

We employed the tokenizer from the DistilBERT model \citep{sanh2019distilbert} to
convert source codes into sequences of numerical tokens. The tokenization
parameters were configured as follows:

\begin{itemize}
    \item \textbf{Maximum length}: 512 tokens.
    \item \textbf{Truncation}: enabled for exceeding sequences.
    \item \textbf{Padding}: enabled to uniformize sequence length.
    \item \textbf{Attention masks}: automatically generated to distinguish real
    tokens from padding.
\end{itemize}

This integrated tokenization approach leverages the model's pre-trained subword
vocabulary, which provides a more natural representation of programming language
constructs compared to the traditional Deep Learning models like DNNs and CNNs in
our previous work \citep{correia2025discovering}. The subword tokenization
effectively captures common programming patterns while maintaining the ability to
process unseen tokens through their constituent parts.

\subsubsection{Data Splitting}
\label{subsubsec:data_splitting}

To ensure robust evaluation and prevent overfitting, we employed 10-fold
cross-validation using the Scikit-learn
implementation \citep{scikit-learn}. For each fold, the data was subdivided as
follows:

\begin{itemize}
    \item \textbf{Training set}: 64\% of the data.
    \item \textbf{Validation set}: 16\% of the data.
    \item \textbf{Test set}: 20\% of the data.
\end{itemize}

This partitioning was achieved through a two-step process within each fold: first
separating the test set (20\%), then dividing the remaining 80\% into training
(64\%) and validation (16\%) subsets. This approach maintained consistent class
distributions across all splits while preserving the independence of the test set
throughout the cross-validation process.

\subsection{Model Setup and Training}
\label{subsec:model_training}

\subsubsection{Model Selection}
\label{subsubsec:model_selection}

We selected the \textbf{DistilBERT} model, a lighter and more efficient
Transformer architecture that maintains advanced contextual understanding
capabilities. This choice was based on the following criteria:

\begin{itemize}
    \item \textbf{Computational efficiency}: distilled version of BERT, suitable
    for iterative experiments.
    \item \textbf{Context capture}: ability to understand complex relationships in
    structured language, such as source code.
    \item \textbf{Direct processing}: eliminates the need for complex
    preprocessing and manual feature engineering.
    \item \textbf{Pre-training}: leverages linguistic knowledge acquired during
    pre-training on large corpora.
\end{itemize}

The model was loaded from the Hugging Face Transformers
library \citep{wolf2020transformers} using the \texttt{distilbert-base-uncased}
version.

\subsubsection{Architecture and Hyperparameters}
\label{subsubsec:architecture_hyperparameters}

We configured the model for binary classification with the following
specifications:

\begin{itemize}
    \item \textbf{Number of labels}: 2 (Parallelizable, Undefined).
    \item \textbf{Number of epochs}: 50.
    \item \textbf{Batch size}: 16.
    \item \textbf{Optimizer}: AdamW \citep{loshchilov2017decoupled}.
    \item \textbf{Learning rate}: $2 \times 10^{-5}$.
    \item \textbf{Learning rate scheduler}: linear with warmup.
    \item \textbf{Loss function}: Binary Cross-Entropy \citep{goodfellow2016deep}.
    \item \textbf{Early stopping}: patience of 50 epochs.
\end{itemize}

These hyperparameters were defined after a series of preliminary tests with
different configurations, where we explored various parameter combinations. The
selected configuration proved to be the most suitable for the task, demonstrating
stability during training and good generalization performance, with no evidence of
overfitting or convergence issues.

\subsubsection{Training Process}
\label{subsubsec:training_process}

The training was conducted using the \texttt{Trainer} class from the Transformers
library, with the following configurations:

\begin{itemize}
    \item \textbf{Evaluation strategy}: every epoch.
    \item \textbf{Saving strategy}: every epoch.
    \item \textbf{Metric for best model}: lowest validation loss.
    \item \textbf{Loading best model at end}: enabled.
    \item \textbf{Save limit}: 1 checkpoint.
    \item \textbf{Early stopping callback}: \texttt{EarlyStoppingCallback} with
    patience of 50 epochs.
\end{itemize}

For each of the 10 folds, the model was reinitialized and trained from scratch,
ensuring independence between executions. The early stopping callback was
configured with a patience of 50 epochs, serving as a contingency mechanism while
allowing complete training cycles. The best model for each fold was automatically
selected based on validation loss and loaded at the end of training.

\subsection{Experimental Evaluation}
\label{subsec:evaluation}

\subsubsection{Performance Metrics}
\label{subsubsec:metrics}

We evaluated the model using multiple metrics to obtain a comprehensive
performance overview:

\begin{itemize}
    \item \textbf{Accuracy}: proportion of correct classifications.
    \item \textbf{Precision}: ability to avoid classifying non-parallelizable
    loops as parallelizable.
    \item \textbf{Recall}: ability to identify all parallelizable loops.
    \item \textbf{F1-score}: harmonic mean between precision and recall.
    \item \textbf{Confusion matrix}: detailed analysis of classification errors.
\end{itemize}

All metrics were calculated using Scikit-learn functions, with binary averaging
for precision, recall, and F1-score.

\subsubsection{Evaluation Procedures}
\label{subsubsec:evaluation_procedures}

For each of the 10 folds, we performed the following analyses:

\begin{itemize}
    \item \textbf{Learning curves}: plots of loss and accuracy during training.
    \item \textbf{Per-class metrics}: precision, recall, and F1-score for each
    class.
    \item \textbf{Confusion matrices}: analysis of error patterns.
    \item \textbf{Cross-fold comparison}: stability and consistency of performance
    across folds.
\end{itemize}

The evaluation included comprehensive analysis of all 10 trained models, with
detailed comparison of training, validation, and test performance for each fold.

\subsubsection{Statistical Analysis}
\label{subsubsec:statistical_analysis}

Data analysis was performed through a dedicated Python workflow that integrated
specialized libraries for the calculation of descriptive statistics and confidence
intervals. For each metric---training accuracy, validation accuracy, and test
accuracy---we computed the mean, sample standard deviation, median, and the
\textbf{95\% confidence interval} using the Student's \textit{t}-distribution
over the sample means \citep{student1908probable}. These metrics were obtained
from the ten executions corresponding to the ten folds used in the experimental
procedure, ensuring consistency with the overall evaluation protocol.

To visually characterize the variability of the results, \textbf{boxplots} were
generated for all accuracy metrics, including the explicit representation of
potential outliers when present. Complementarily, \textbf{histograms} were
plotted to examine the distribution shape of the accuracy values across the folds.

Beyond accuracy, the analysis also included the mean, standard deviation, median,
and 95\% confidence intervals for the \textbf{F1-score, precision, and recall},
computed over the ten folds using the same Student's
\textit{t}-distribution approach \citep{student1908probable}.

A detailed analysis of the \textbf{validation error} was also performed across
all folds. For each fold, we computed the mean, standard deviation, median, and
the 95\% confidence interval of the validation loss, measured using the binary
cross-entropy function. This evaluation was conducted because the \textbf{validation
error constitutes one of the key criteria} used to identify the best and
worst-performing models---specifically, the model exhibiting the \textit{lowest}
validation error.

In an analogous manner, a comprehensive statistical analysis of the
\textbf{false positive rate (FPR)} was conducted across all folds. For this
metric, we computed the mean, standard deviation, median, and the
\textbf{95\% confidence interval} using the same Student's
\textit{t}-distribution procedure applied to the validation error. This analysis
enables a robust assessment of the model's tendency to incorrectly classify
instances from the Undefined class (negative) as Parallelizable (positive), which
represents a critical error in the context of automatic parallelization.

Following this analysis, we proceed to the identification of the \textbf{best}
and \textbf{worst} cases according to the two selection criteria adopted in this
study. The first criterion selects the model with the lowest validation error
(best case) and the model with the highest validation error (worst case). The
second criterion focuses on the model's behavior regarding the \textbf{critical
error} for our application: the \textit{false positive rate (FPR)}, defined as
the proportion of instances from the Undefined class (negative) that are
incorrectly predicted as Parallelizable (positive). Under this criterion, the
best model is the one that presents a zero or minimal \textbf{FPR}, while the
worst model is the one exhibiting the highest \textbf{FPR}.

For each selected model---best and worst under each criterion---we present the
results in the following order: (1) the training and validation curves,
(2) the confusion matrix, and (3) the classification report. This structured
evaluation enables a thorough inspection of model behavior under both
performance-oriented and application-critical perspectives.

\section{Results}
\label{sec:results}

This section provides a comprehensive analysis of the performance of the
DistilBERT model in classifying parallelizable and undefined loops. The results
are organized progressively, beginning with an overview of the model's average
behavior across all folds, followed by a detailed analysis of accuracy.
Subsequent subsections present the remaining performance metrics---precision,
recall, and F1-score---in a dedicated evaluation. The section also incorporates
two independent criteria for selecting representative models: the validation error
and the number of false positives. Because these criteria assess different and
equally relevant aspects of performance, four models are examined in the best-
and worst-case analyses: the model with the lowest validation error and the one
with the lowest number of false positives (best cases), as well as the model with
the highest validation error and the one with the highest number of false
positives (worst cases). The section concludes with a discussion of the results
and the model's robustness. All evaluations were conducted using 10-fold
cross-validation, ensuring statistically robust and reliable analysis.

\subsection{Average Model Behavior}
\label{subsec:average_behavior}

\begin{figure}[htbp]
\centering
\includegraphics[width=0.6\linewidth]{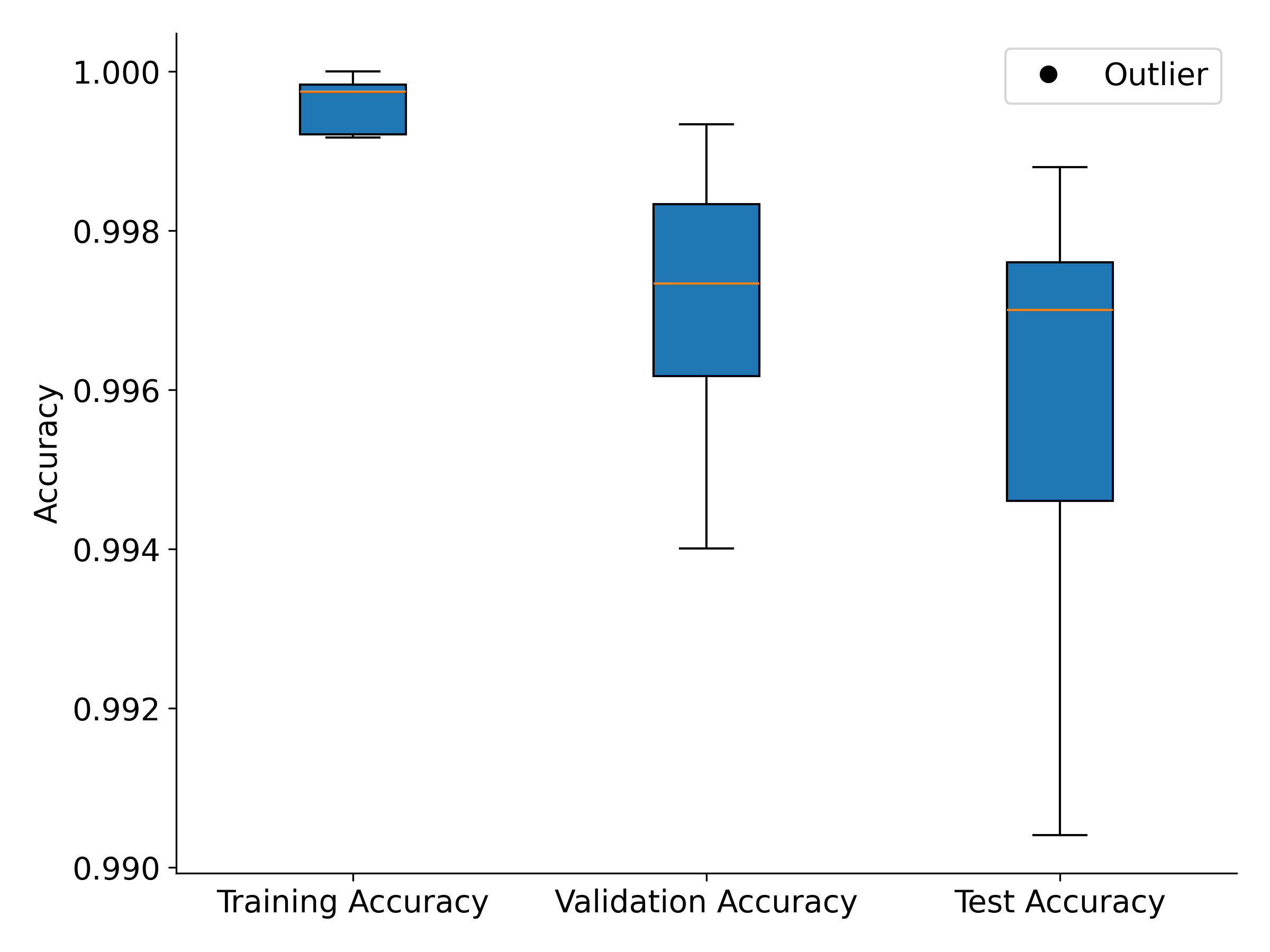}
\caption{Boxplot of training, validation, and test accuracies for the 10
cross-validation folds.}
\label{fig:boxplot_accuracies}
\end{figure}

\begin{figure}[htbp]
\centering
\includegraphics[width=0.6\linewidth]{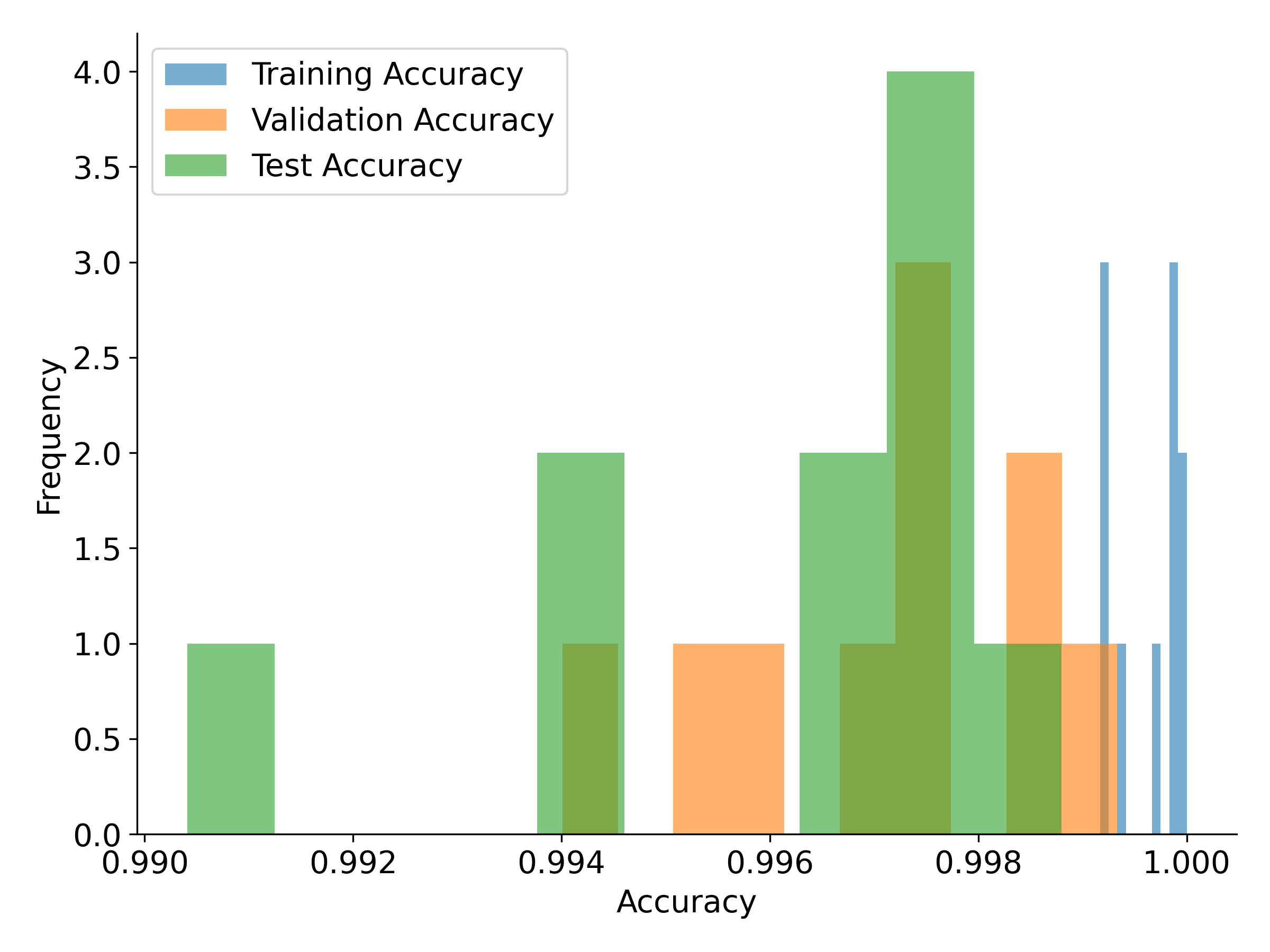}
\caption{Histograms of training, validation, and test accuracies for the 10
folds.}
\label{fig:histogram_accuracies}
\end{figure}

To assess the average behavior of the DistilBERT model across all
cross-validation folds, we analyzed the accuracy distributions for the training,
validation, and test sets. Figure~\ref{fig:boxplot_accuracies} presents a boxplot
that statistically summarizes these distributions, while
Figure~\ref{fig:histogram_accuracies} shows the corresponding histograms,
providing a more detailed visualization of accuracy frequency.

The results show that the model achieved high and highly consistent performance
across all evaluated sets. The mean training accuracy was approximately 0.99965,
indicating excellent ability to fit the data. The mean validation accuracy, around
0.99753, demonstrates that the model maintains very high performance even on
unseen data, with low variation across folds. The mean test accuracy, equal to
0.99600, reinforces the model's robustness and the absence of overfitting, as
validation and test performances remain close and at high levels. The small
dispersion observed in both the boxplot and histograms highlights the stability of
DistilBERT for the proposed task.

\subsubsection{Overall Accuracy Performance}
\label{subsubsec:accuracy_performance}

For a more detailed quantitative analysis of the model's performance,
Table~\ref{tab:accuracy_statistics} presents the main consolidated statistical
metrics of accuracies obtained in the training, validation, and test sets across
the 10 cross-validation folds.

\begin{table}[htbp]
\centering
\caption{Statistical metrics for training, validation, and test accuracies across
the 10 cross-validation folds.}
\label{tab:accuracy_statistics}
\begin{tabular}{lcccc}
\toprule
\textbf{Set} & \textbf{Mean} & \textbf{Std.\ Dev.} & \textbf{Median} & \textbf{95\% CI} \\
\midrule
Training   & 0.99965 & 0.00030 & 0.99983 & 0.99944--0.99986 \\
Validation & 0.99753 & 0.00109 & 0.99784 & 0.99675--0.99831 \\
Test       & 0.99600 & 0.00261 & 0.99650 & 0.99413--0.99787 \\
\bottomrule
\end{tabular}
\end{table}

The results demonstrate exceptionally high and consistent performance across all
sets. The mean training accuracy of 99.96\% confirms the model's ability to
effectively learn the patterns present in the data. The mean validation (99.75\%)
and test (99.60\%) accuracies show excellent generalization capability, maintaining
very close and high-level results.

The analysis of the standard deviation reveals low variability across folds,
reinforcing the model's stability. The training set exhibited the lowest
variability (SD = 0.00030), followed by validation (SD = 0.00109) and test
(SD = 0.00261). The proximity between means and medians indicates balanced
distributions with no relevant skewness.

The 95\% confidence intervals, calculated using Student's $t$-test, provide a
robust estimate of mean precision. For the test set, for example, we can state
with 95\% confidence that the mean accuracy lies between 99.41\% and 99.78\%. For
validation, the interval ranges between 99.67\% and 99.83\%. These narrow
intervals further reinforce the strong statistical reliability of the results.

The combination of high mean accuracies, low variability, and narrow confidence
intervals confirms the robustness of DistilBERT for the classification task,
showing that the model consistently maintains its performance when exposed to new
data.

\subsubsection{Performance on Precision, Recall, and F1-Score}
\label{subsubsec:detailed_metrics}

\begin{figure}[htbp]
\centering
\includegraphics[width=0.6\linewidth]{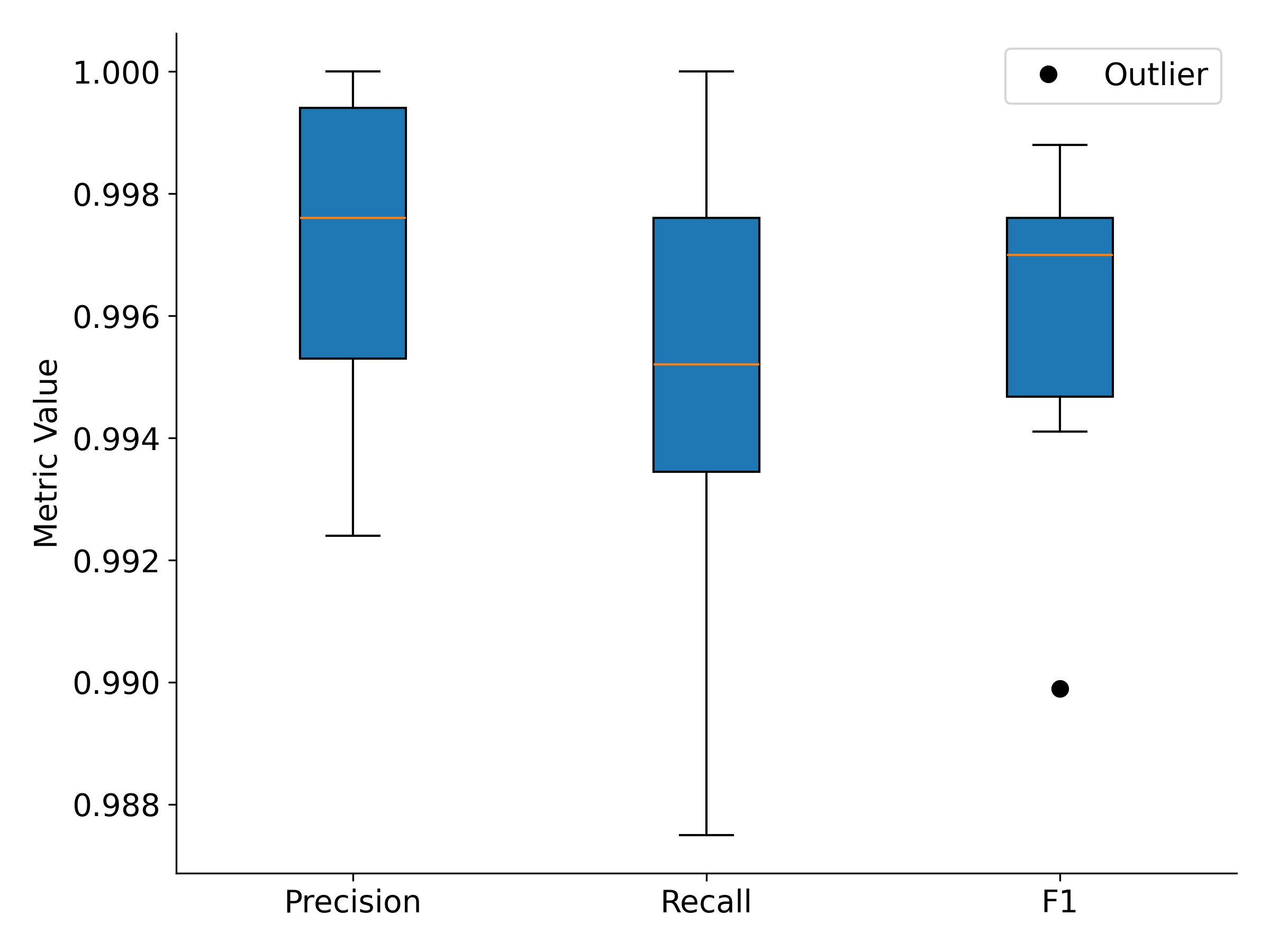}
\caption{Boxplots of precision, recall, and F1-Score metrics across the 10
cross-validation folds.}
\label{fig:boxplot_all_metrics}
\end{figure}

\begin{figure}[htbp]
\centering
\includegraphics[width=0.6\linewidth]{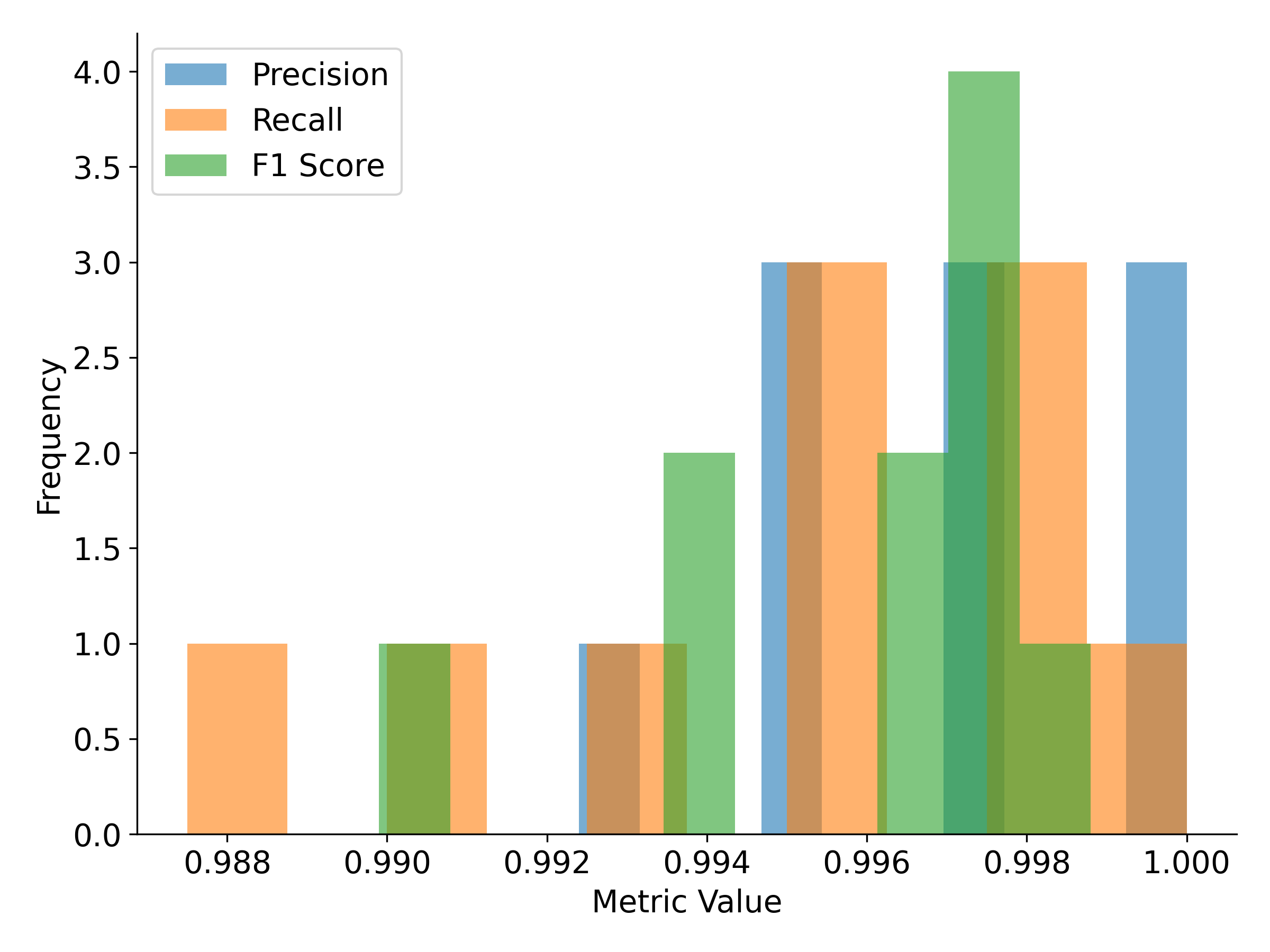}
\caption{Histograms of precision, recall, and F1-Score metrics across the 10
cross-validation folds.}
\label{fig:histogram_all_metrics}
\end{figure}

To complement the analysis of the model's performance, we evaluated the
precision, recall, and F1-Score metrics on the test set.
Figure~\ref{fig:boxplot_all_metrics} presents the distribution of these metrics
using boxplots, while Figure~\ref{fig:histogram_all_metrics} shows the
corresponding histograms, allowing for a detailed visual analysis of the behavior
of the three metrics across the 10 cross-validation folds.

The visualizations show that all metrics maintain consistently high values, with
most observations concentrated above 0.990. Recall exhibits a slightly more
compact distribution, indicating stable performance in correctly identifying
positive examples. Precision, although exhibiting slightly greater variation
across folds, remains consistently high in all cases. The F1-Score, as a balanced
metric between precision and recall, reflects an intermediate behavior while
maintaining high stability across the folds.

For a more detailed quantitative analysis, Table~\ref{tab:detailed_metrics}
presents the consolidated statistics for the three metrics:

\begin{table}[htbp]
\centering
\caption{Statistical metrics for Precision, Recall, and F1-Score on the test
set.}
\label{tab:detailed_metrics}
\begin{tabular}{lcccc}
\toprule
\textbf{Metric} & \textbf{Mean} & \textbf{Std.\ Dev.} & \textbf{Median} & \textbf{95\% CI} \\
\midrule
Precision & 0.9971 & 0.0026 & 0.9976 & 0.9953--0.9989 \\
Recall    & 0.9949 & 0.0038 & 0.9952 & 0.9922--0.9976 \\
F1-Score  & 0.9960 & 0.0026 & 0.9970 & 0.9941--0.9979 \\
\bottomrule
\end{tabular}
\end{table}

The numerical results confirm the visual observations from the figures. Precision
exhibits the highest mean among the metrics (99.71\%) with low variability
(SD = 0.0026), indicating stability in the model's ability to avoid false
positives. Recall, with a mean of 99.49\%, also demonstrates strong performance
in identifying true positives, although with slightly higher variability
(SD = 0.0038). The F1-Score, with a mean of 99.60\%, reflects an appropriate
balance between the two previous metrics, capturing both the model's ability to
correctly identify positive examples and its ability to minimize classification
errors.

The 95\% confidence intervals, computed using Student's $t$-test, reinforce the
reliability of the results: for the F1-Score, the population mean lies between
99.41\% and 99.79\%; for recall, between 99.22\% and 99.76\%; and for precision,
between 99.53\% and 99.89\%. The proximity between means and medians across all
metrics indicates symmetrical and consistent distributions across the folds,
corroborating the model's stability.

The high and consistent performance observed across all evaluated
metrics---precision, recall, and F1-Score---confirms the effectiveness of the
DistilBERT model for the loop classification task. The results demonstrate not
only high overall accuracy but also a proper balance between the ability to
correctly identify positive examples and avoid false positives, reinforcing the
robustness of the model.

\subsection{Analysis of the Validation Error as a Selection Criterion}
\label{subsec:validation_error}

The criterion used to select the best model in each fold was based on the lowest
validation error, employing the \textit{Binary Cross-Entropy} function as the
evaluation metric during training. This approach makes it possible to identify the
points of best model generalization, avoiding overfitting and ensuring that
performance is preserved on unseen data.

\begin{figure}[htbp]
\centering
\includegraphics[width=0.6\linewidth]{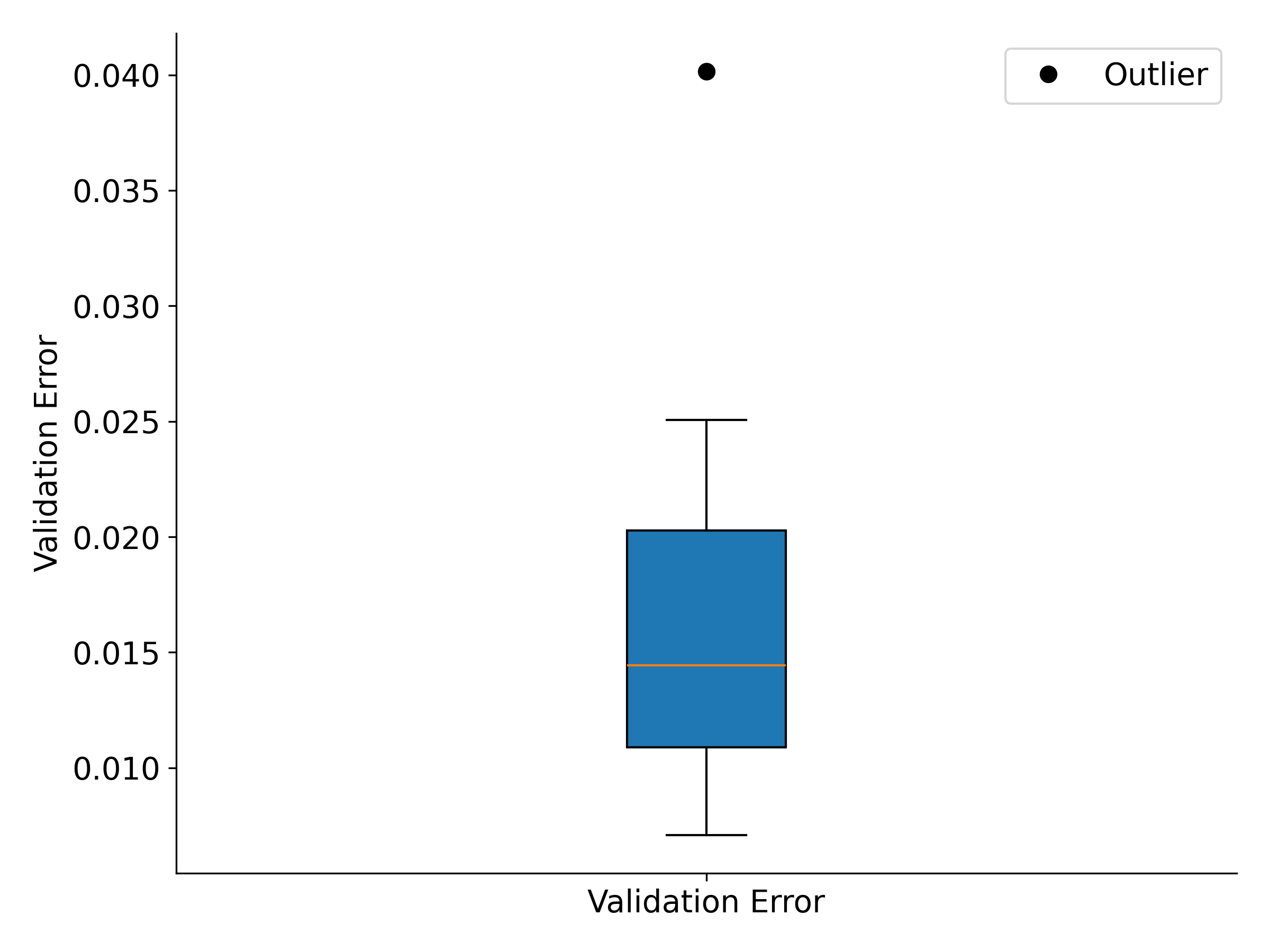}
\caption{Boxplot of the validation error (Binary Cross-Entropy) for the 10
folds.}
\label{fig:boxplot_eval_loss}
\end{figure}

\begin{figure}[htbp]
\centering
\includegraphics[width=0.6\linewidth]{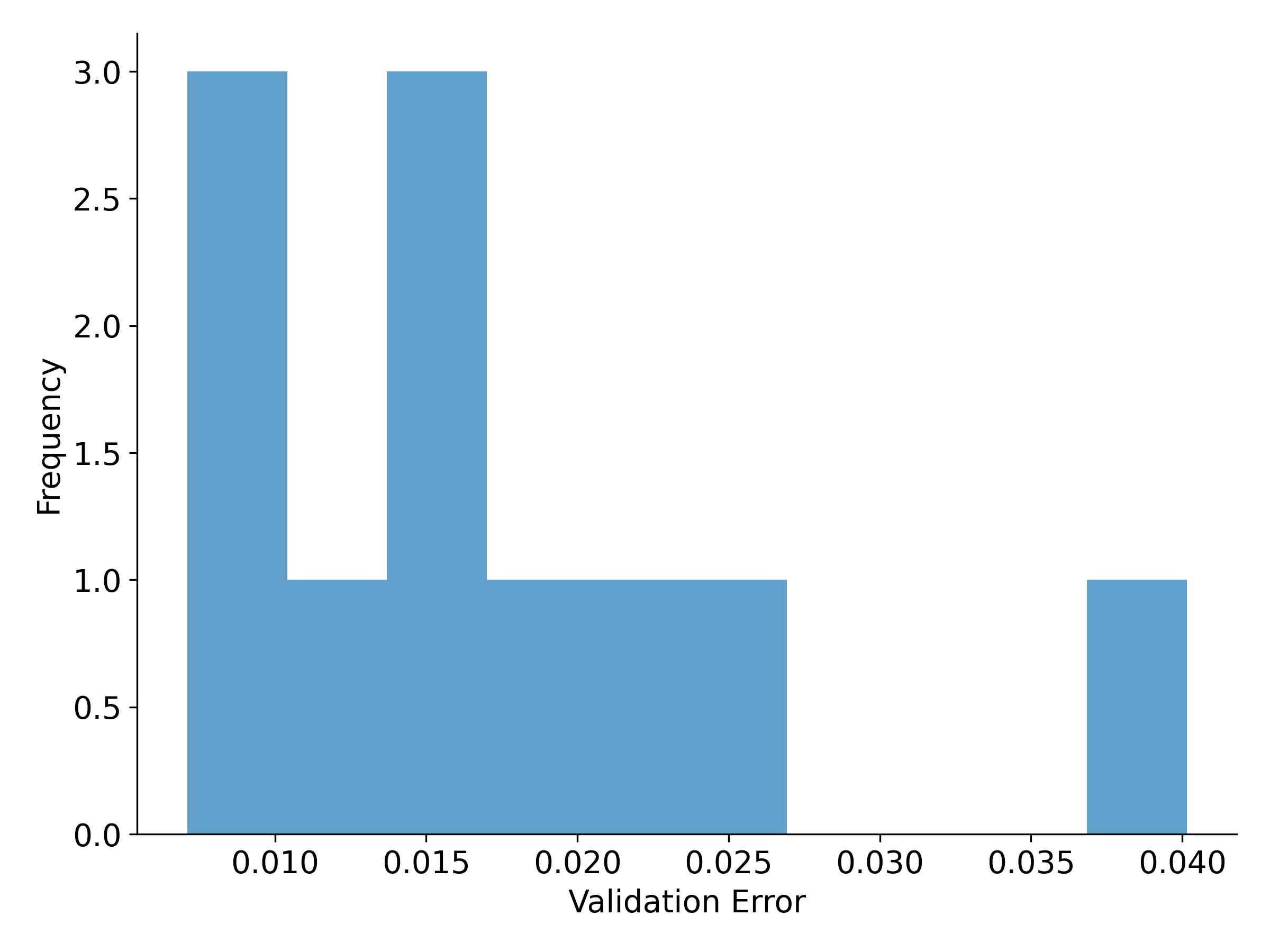}
\caption{Histogram of the validation error (Binary Cross-Entropy) for the 10
folds.}
\label{fig:histogram_eval_loss}
\end{figure}

Figure~\ref{fig:boxplot_eval_loss} presents the distribution of the validation
error through a boxplot, while Figure~\ref{fig:histogram_eval_loss} shows the
corresponding histogram, allowing a detailed visual inspection of the error
behavior across the 10 folds.

The visualizations show that the validation error exhibits consistently low
values, ranging approximately from 0.007 to 0.040. The boxplot reveals a compact
distribution, with most values concentrated in the lower range, indicating good
stability across the folds. The histogram reveals an asymmetric distribution with
higher density in the lowest error values, which is desirable for an error metric.

For a more detailed quantitative analysis, Table~\ref{tab:validation_error_stats}
presents the consolidated statistics of the validation error:

\begin{table}[htbp]
\centering
\caption{Statistical metrics of the validation error (Binary Cross-Entropy).}
\label{tab:validation_error_stats}
\begin{tabular}{lcccc}
\toprule
\textbf{Metric} & \textbf{Mean} & \textbf{Std.\ Dev.} & \textbf{Median} & \textbf{95\% CI} \\
\midrule
Validation Error & 0.01772 & 0.00965 & 0.01512 & 0.01081--0.02462 \\
\bottomrule
\end{tabular}
\end{table}

The numerical results confirm the visual observations. The mean validation error
of 0.01772, combined with the median of 0.01512, indicates that most folds
exhibit very low error values, with a few slightly higher values pulling the mean
upward---as reflected by the maximum of approximately 0.040. The standard
deviation of 0.00965 reveals moderate variability, also reflected in the extent
of the whiskers in the boxplot.

The 95\% confidence interval, estimated between 0.01081 and 0.02462, shows that
the population mean validation error is concentrated within a narrow range,
reinforcing the stability of the validation process. The combination of low
absolute error values and the consistency observed across the folds validates the
effectiveness of the selection criterion based on the lowest validation error,
demonstrating that this approach is suitable for identifying models with strong
generalization capability in the proposed classification task.

\subsection{False Positive Rate (FPR) Analysis}
\label{subsec:fpr_analysis}

In addition to using the validation error as the selection criterion for the
best-performing model, we also conducted a detailed analysis of the False Positive
Rate (FPR) across all folds of the cross-validation. The FPR represents the
proportion of negative instances incorrectly classified as positive, which is
particularly relevant in software defect prediction tasks. A high FPR would lead
to false alarms, unnecessarily increasing the inspection effort and causing
inefficient allocation of maintenance resources. Therefore, examining the
statistical behavior of this metric provides further insight into the reliability
and robustness of the model.

\begin{figure}[htbp]
    \centering
    \includegraphics[width=0.55\linewidth]{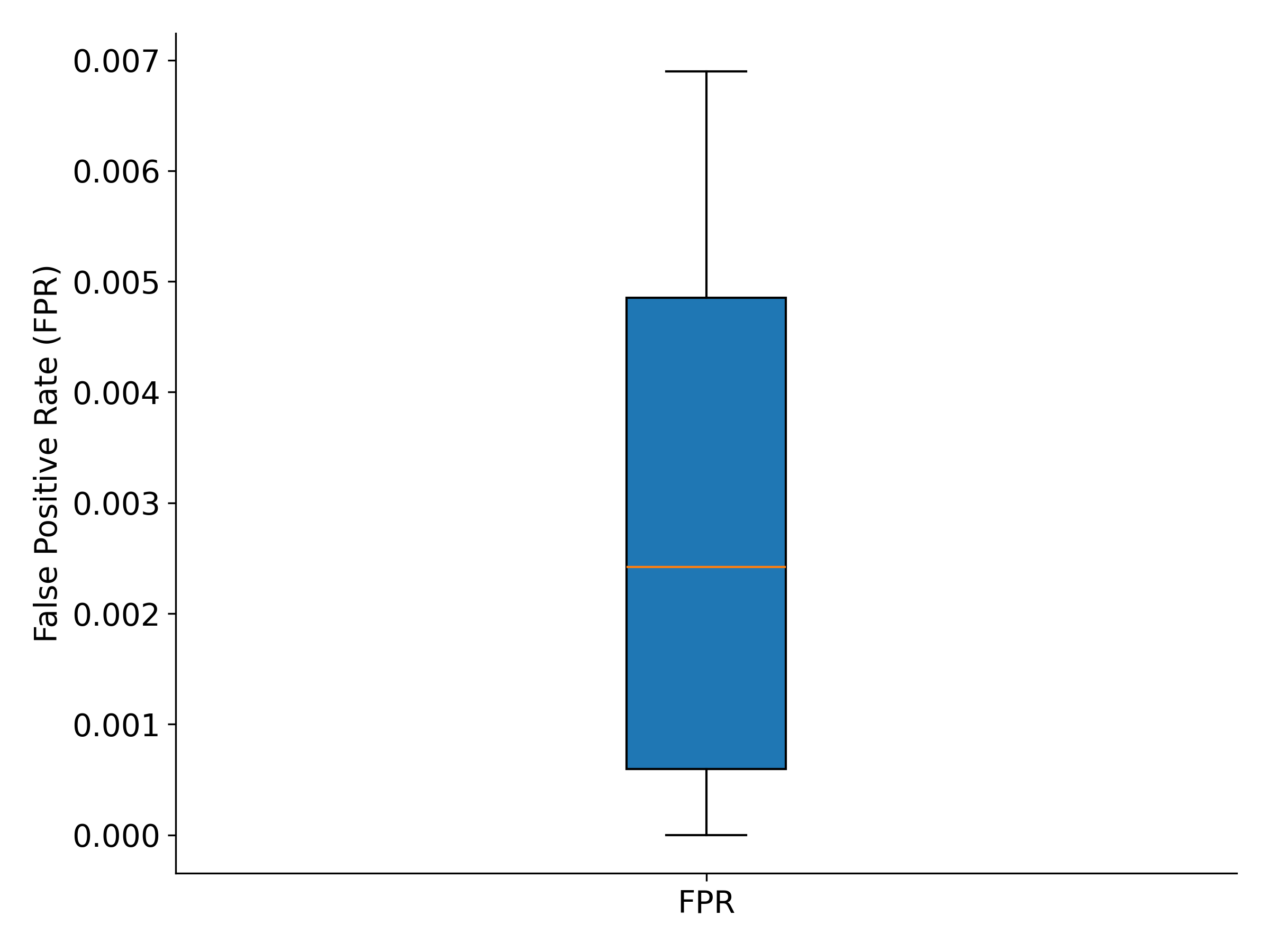}
    \caption{Boxplot of the False Positive Rate (FPR) across the 10-fold
    cross-validation.}
    \label{fig:fpr_boxplot}
\end{figure}

\begin{figure}[htbp]
    \centering
    \includegraphics[width=0.55\linewidth]{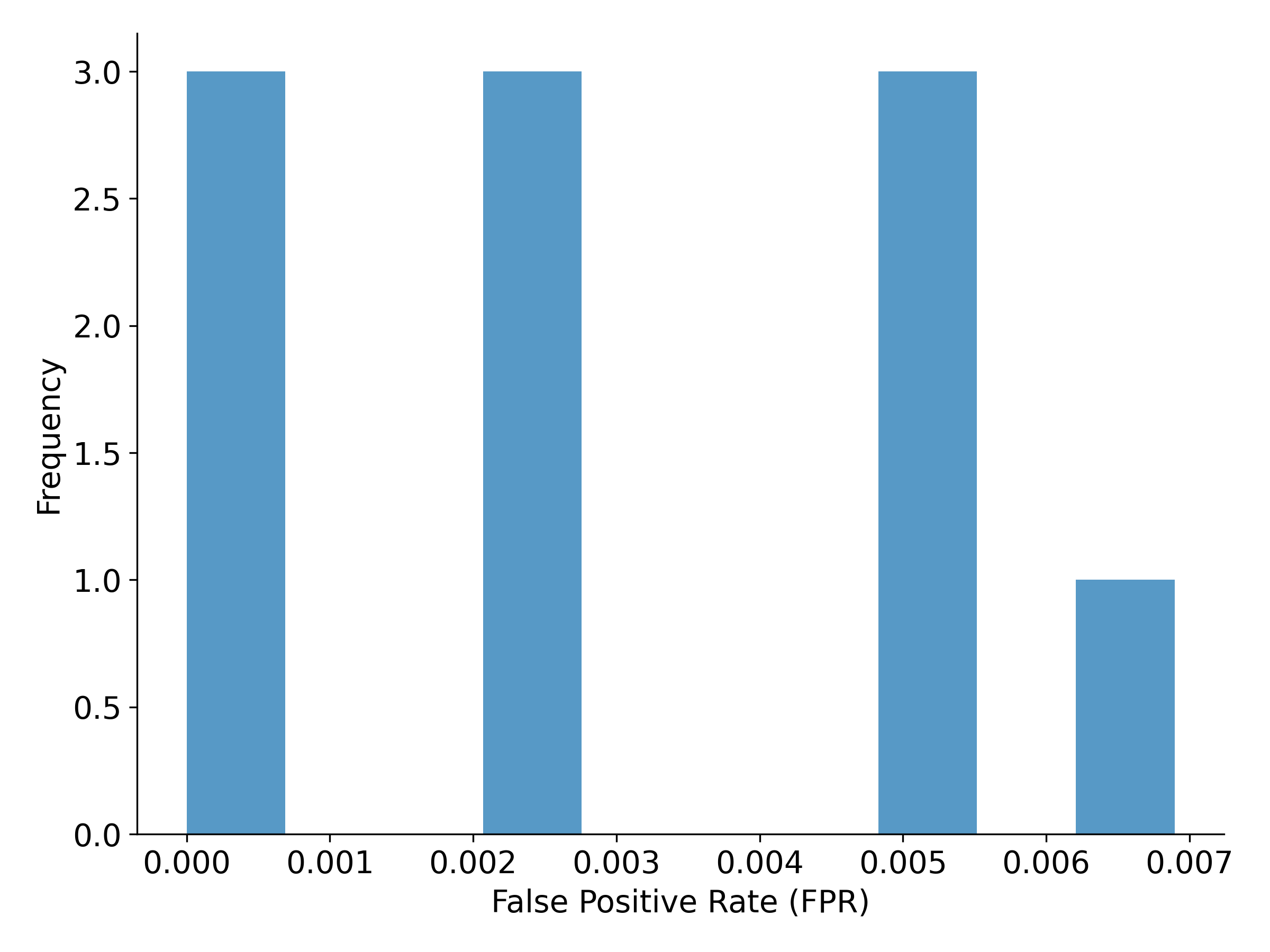}
    \caption{Histogram of the False Positive Rate (FPR) across the 10-fold
    cross-validation.}
    \label{fig:fpr_histogram}
\end{figure}

Figure~\ref{fig:fpr_boxplot} presents the boxplot of the FPR values. The
distribution is highly concentrated near zero, reflecting that the model rarely
produces false alarms. Only a few folds exhibit small non-zero values, and even
in these cases, the rate remains extremely low. Figure~\ref{fig:fpr_histogram}
shows the corresponding histogram, reinforcing the observation that most folds
lie within the lowest bins of the distribution.

\begin{table}[htbp]
\centering
\caption{Statistical metrics for the False Positive Rate (FPR) across the 10-fold
cross-validation.}
\label{tab:fpr_statistics}
\begin{tabular}{lcccc}
\toprule
\textbf{Metric} & \textbf{Mean} & \textbf{Std.\ Dev.} & \textbf{Median} & \textbf{95\% CI} \\
\midrule
FPR & 0.002872 & 0.002439 & 0.002418 & 0.001197--0.004546 \\
\bottomrule
\end{tabular}
\end{table}

Table~\ref{tab:fpr_statistics} summarizes the descriptive statistics for the FPR
across the 10 folds. The mean FPR is 0.002872, and the median (0.002418) is of
the same magnitude, indicating consistent behavior among the folds. The standard
deviation is low, demonstrating limited variability, and the 95\% confidence
interval (0.001197--0.004546) remains tightly bounded, further confirming the
stability of the model regarding false positive generation.

\subsection{Analysis of the Best-Performing Models}
\label{subsec:best_models}

Based on the two selection criteria previously discussed---the lowest validation
error and the lowest critical error (False Positive Rate)---this section presents
a detailed analysis of the best-performing models identified during the 10-fold
cross-validation process. For each criterion, the selected model is analyzed
through its training dynamics, confusion matrix, and classification metrics,
providing a comprehensive evaluation of its performance and reliability.

\subsubsection{Best Model According to the Lowest Validation Error}
\label{subsubsec:best_val_error}

Among the ten folds evaluated, \textbf{Fold 7} achieved the lowest validation
error, making it the best-performing model according to the validation loss
criterion. This result indicates superior generalization capability, as the model
minimizes the Binary Cross-Entropy loss on unseen validation data.

\begin{figure}[htbp]
\centering
\includegraphics[width=0.95\linewidth]{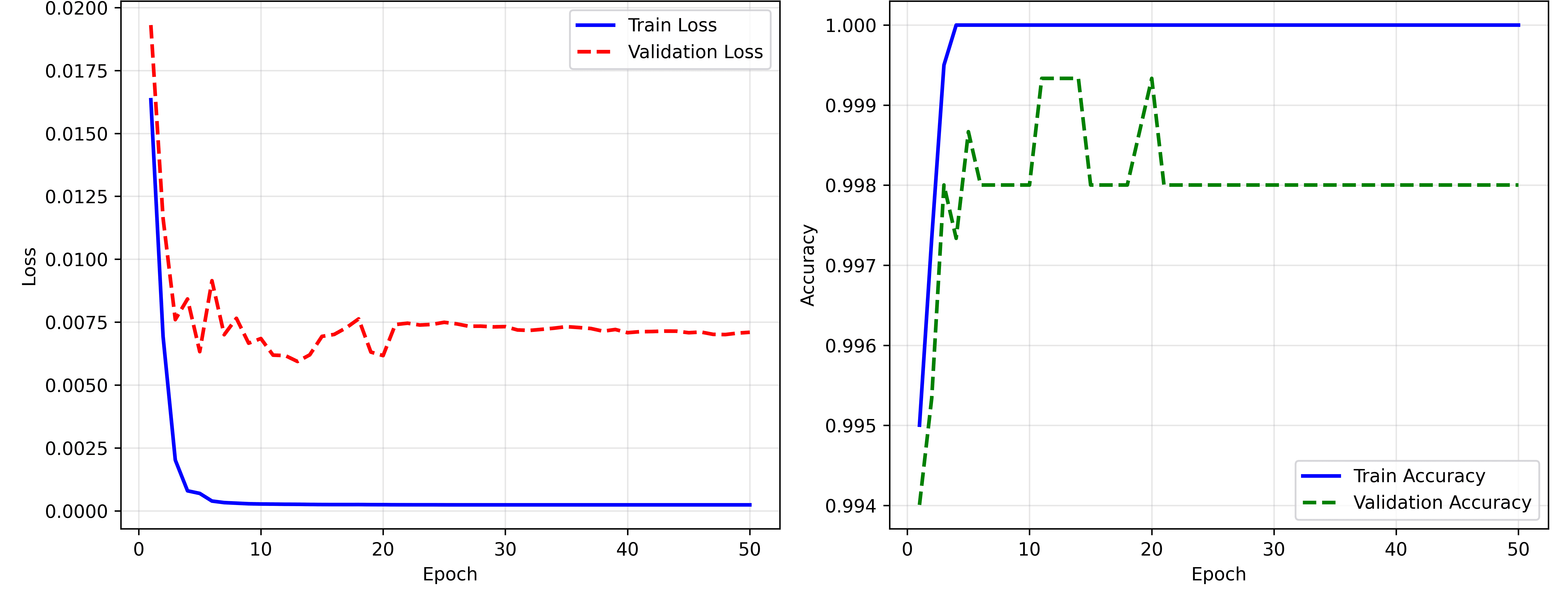}
\caption{Training and validation loss and accuracy curves for Fold 7.}
\label{fig:training_fold7}
\end{figure}

Figure~\ref{fig:training_fold7} presents the training and validation loss curves,
along with the corresponding accuracy curves. The training loss rapidly converges
toward zero, while the validation loss stabilizes at a low value, showing no signs
of divergence or instability. Additionally, the training accuracy reaches 100\%,
and the validation accuracy remains consistently high throughout the training
process, indicating a well-fitted model without evidence of overfitting.

The classification performance of Fold 7 is summarized by the confusion matrix
presented in Table~\ref{tab:confusion_fold7}. The results show that the model
correctly classified almost all instances, with only a minimal number of
misclassifications.

\begin{table}[htbp]
\centering
\caption{Confusion matrix for Fold 7.}
\label{tab:confusion_fold7}
\begin{tabular}{lcc}
\toprule
\textbf{Actual / Predicted} & \textbf{Undefined} & \textbf{Independent} \\
\midrule
Undefined    & 414 & 1 \\
Independent  & 2   & 417 \\
\bottomrule
\end{tabular}
\end{table}

Quantitatively, Fold 7 achieved 417 true positives (TP), 414 true negatives (TN),
only 2 false negatives (FN), and 1 false positive (FP).
Table~\ref{tab:classification_fold7} presents the corresponding classification
metrics. The model achieved a recall of 0.9952, precision of 0.9976, and an
F1-score of 0.9964, demonstrating an excellent balance between sensitivity and
precision.

\begin{table}[htbp]
\centering
\caption{Classification report for Fold 7.}
\label{tab:classification_fold7}
\begin{tabular}{lccc}
\toprule
\textbf{Metric} & \textbf{Precision} & \textbf{Recall} & \textbf{F1-score} \\
\midrule
Independent Class & 0.9976 & 0.9952 & 0.9964 \\
\bottomrule
\end{tabular}
\end{table}

\subsubsection{Best Model According to the Lowest Critical Error}
\label{subsubsec:best_fpr}

Considering the False Positive Rate (FPR) as the critical error criterion,
\textbf{Fold 8} emerged as the best-performing model. This fold achieved the
lowest FPR among all evaluated models, which is particularly relevant for the
proposed task, as false positives may lead to unnecessary inspections and
inefficient allocation of maintenance resources.

\begin{figure}[htbp]
\centering
\includegraphics[width=0.95\linewidth]{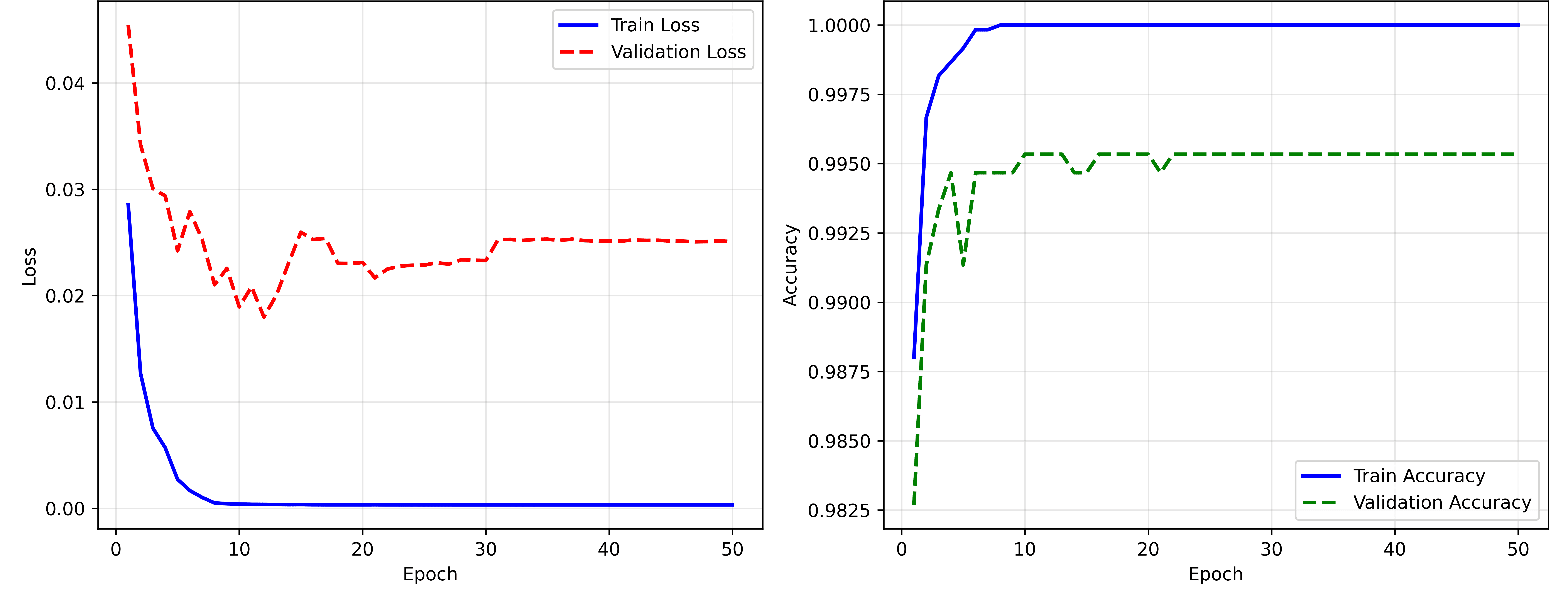}
\caption{Training and validation loss and accuracy curves for Fold 8.}
\label{fig:training_fold8}
\end{figure}

Figure~\ref{fig:training_fold8} presents the training and validation curves for
Fold 8. The training process shows fast convergence, with training loss approaching
zero and validation loss stabilizing at a low level. The training accuracy reaches
100\%, while the validation accuracy remains consistently high, indicating stable
learning behavior and strong generalization.

The confusion matrix for Fold 8 is presented in Table~\ref{tab:confusion_fold8}.
Notably, this model produced no false positives, which directly explains its
selection under the critical error criterion.

\begin{table}[htbp]
\centering
\caption{Confusion matrix for Fold 8.}
\label{tab:confusion_fold8}
\begin{tabular}{lcc}
\toprule
\textbf{Actual / Predicted} & \textbf{Undefined} & \textbf{Independent} \\
\midrule
Undefined    & 419 & 0 \\
Independent  & 1   & 414 \\
\bottomrule
\end{tabular}
\end{table}

From a quantitative perspective, Fold 8 achieved 414 true positives (TP), 419
true negatives (TN), 1 false negative (FN), and zero false positives (FP).
Table~\ref{tab:classification_fold8} summarizes the classification metrics. The
model attained a recall of 0.9976, perfect precision of 1.0000, and an F1-score
of 0.9988, highlighting its robustness in minimizing critical classification
errors.

\begin{table}[htbp]
\centering
\caption{Classification report for Fold 8.}
\label{tab:classification_fold8}
\begin{tabular}{lccc}
\toprule
\textbf{Metric} & \textbf{Precision} & \textbf{Recall} & \textbf{F1-score} \\
\midrule
Independent Class & 1.0000 & 0.9976 & 0.9988 \\
\bottomrule
\end{tabular}
\end{table}

Overall, the analysis of both selected models demonstrates that the proposed
approach is capable of producing highly accurate and reliable classifiers under
different optimization criteria. While Fold 7 excels in terms of overall
generalization as indicated by the lowest validation error, Fold 8 provides
superior performance in minimizing critical errors, offering a complementary
perspective for model selection depending on application-specific requirements.

\subsection{Analysis of the Worst-Performing Models}
\label{subsec:worst_models}

Although the proposed approach demonstrated highly stable performance across all
folds, an analysis of the worst-performing cases provides additional insight into
the lower bounds of the method under different selection criteria. Using the same
criteria adopted for the best-performing models---highest validation error and
highest critical error (False Positive Rate)---this section discusses the least
favorable scenarios observed during the cross-validation process.

\subsubsection{Worst Model According to the Highest Validation Error}
\label{subsubsec:worst_val_error}

With respect to validation loss, \textbf{Fold 8} presented the highest validation
error among all evaluated folds. This behavior indicates comparatively weaker
generalization performance according to the Binary Cross-Entropy objective.

It is important to note that Fold 8 was previously analyzed in detail in
Section~\ref{subsubsec:best_fpr}, where it was identified as the best-performing
model under the critical error criterion. Therefore, its training dynamics,
confusion matrix, and classification report are not repeated here to avoid
redundancy.

From a quantitative perspective, Fold 8 achieved 414 true positives (TP), 419
true negatives (TN), 1 false negative (FN), and no false positives (FP). Despite
exhibiting the highest validation loss, the model maintains excellent
classification performance, with a recall of 0.9976, perfect precision of 1.0000,
and an F1-score of 0.9988. These results highlight a clear trade-off between
optimization objectives, where minimizing validation error and minimizing critical
errors may lead to different model selections.

\subsubsection{Worst Model According to the Highest Critical Error}
\label{subsubsec:worst_fpr}

Considering the False Positive Rate as the critical error criterion,
\textbf{Fold 2} was identified as the worst-performing model, presenting the
highest number of false positives among all folds. Although the absolute number of
critical errors remains limited, this fold represents the least favorable case in
scenarios where false alarms must be strictly minimized.

\begin{figure}[htbp]
\centering
\includegraphics[width=0.95\linewidth]{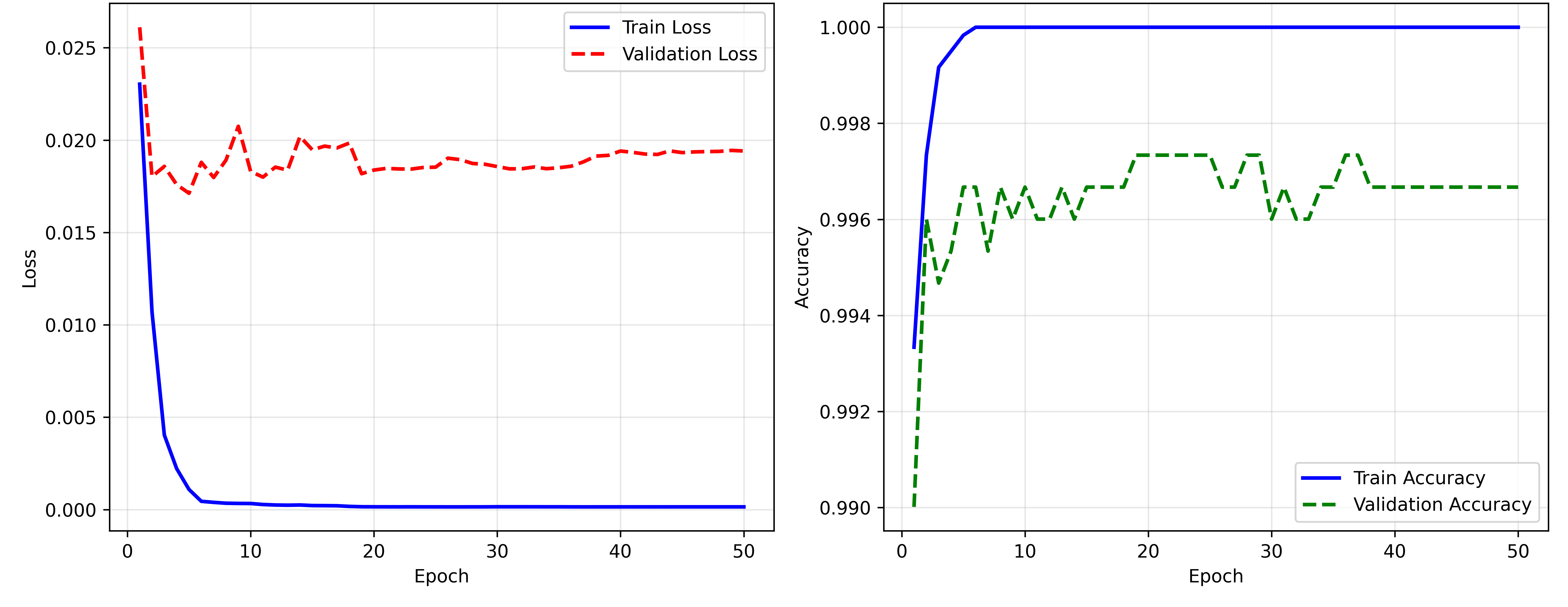}
\caption{Training and validation loss and accuracy curves for Fold 2 (worst
critical error).}
\label{fig:training_worst_fold2}
\end{figure}

Figure~\ref{fig:training_worst_fold2} illustrates the training and validation loss
and accuracy curves for Fold 2. The training loss rapidly converges toward zero,
while the validation loss stabilizes without signs of divergence. Both training
and validation accuracy remain high throughout the training process, indicating
stable learning behavior despite the increased critical error.

Table~\ref{tab:confusion_fold2_worst} presents the confusion matrix for Fold 2.
The presence of three false positives directly explains its selection as the worst
case under the critical error criterion.

\begin{table}[htbp]
\centering
\caption{Confusion matrix for Fold 2 (worst critical error).}
\label{tab:confusion_fold2_worst}
\begin{tabular}{lcc}
\toprule
\textbf{Actual / Predicted} & \textbf{Undefined} & \textbf{Independent} \\
\midrule
Undefined    & 432 & 3 \\
Independent  & 5   & 394 \\
\bottomrule
\end{tabular}
\end{table}

Table~\ref{tab:metrics_fold2_worst} summarizes the classification metrics for
Fold 2. While precision, recall, and F1-score remain above 0.98, this fold
exhibits the highest False Positive Rate among the evaluated models,
characterizing it as the least favorable case under the critical error
perspective.

\begin{table}[htbp]
\centering
\caption{Classification report for Fold 2 (worst critical error).}
\label{tab:metrics_fold2_worst}
\begin{tabular}{lccc}
\toprule
\textbf{Metric} & \textbf{Precision} & \textbf{Recall} & \textbf{F1-score} \\
\midrule
Independent & 0.9924 & 0.9875 & 0.9899 \\
Undefined   & 0.9886 & 0.9931 & 0.9908 \\
\bottomrule
\end{tabular}
\end{table}

Overall, this analysis confirms that even the worst-performing models maintain
high predictive performance. The observed variations mainly reflect trade-offs
between optimization criteria rather than instability or poor generalization,
reinforcing the robustness of the proposed approach.

\section{Discussion}
\label{sec:discussion}

The experimental results demonstrate that the proposed DistilBERT-based approach
achieves consistently high performance in the task of classifying parallelization
potential in source code. Across the 10-fold cross-validation, the model exhibited
high accuracy on training, validation, and test sets, with narrow confidence
intervals and low variability. The proximity between validation and test accuracies
indicates that the model generalizes well to unseen data, with no evidence of
overfitting, despite the high capacity of the Transformer architecture.

Beyond overall accuracy, the analysis of precision, recall, and F1-score confirms
that the model maintains a balanced classification behavior. Precision remains
consistently high across folds, indicating strong control over false positives,
while recall values also remain above 99\% on average, demonstrating effective
identification of parallelizable loops. The resulting F1-scores reflect this
balance, reinforcing the robustness of the classifier and its suitability for
tasks where both correct detection and error minimization are critical.

The validation error analysis further supports the stability of the training
process. The Binary Cross-Entropy loss exhibits low absolute values and a compact
distribution across folds, confirming that the optimization procedure converges
reliably. However, the comparison between validation error and application-critical
metrics reveals an important trade-off: the model with the lowest validation loss
is not necessarily the one that minimizes false positives. This finding highlights
the relevance of considering multiple evaluation criteria when selecting models for
practical deployment.

The False Positive Rate analysis provides additional insight into the model's
reliability from an application-oriented perspective. The very low average FPR
observed across all folds indicates that the model rarely misclassifies undefined
loops as parallelizable. Notably, one fold achieved zero false positives,
demonstrating that the proposed approach is capable of completely eliminating this
critical error under certain conditions. This behavior is particularly relevant in
scenarios where incorrect parallelization decisions may lead to unsafe or
inefficient execution.

Finally, the examination of best- and worst-performing cases shows that
performance variations are limited and primarily reflect trade-offs between
optimization objectives rather than instability. Even the least favorable models
maintain high accuracy and F1-scores, confirming the robustness of the approach.
Overall, the results indicate that the proposed method effectively leverages
Transformer-based representations for source code analysis, providing accurate,
stable, and reliable classification within the defined scope of the study.

\section{Conclusion}
\label{sec:conclusion}

This work presented a Transformer-based approach for the automatic classification
of parallelization potential in source code, focusing on the identification of
independent loops that can be safely parallelized. The proposed methodology
combines a carefully constructed and balanced dataset---composed of synthetic codes
generated via evolutionary algorithms and manually annotated real-world
examples---with a fine-tuned DistilBERT model capable of processing source code
directly as text. The experimental evaluation was conducted using a rigorous
10-fold cross-validation protocol and a comprehensive set of performance metrics.

The results demonstrate that the proposed approach achieves consistently high
performance across all folds, with strong generalization capability and low
variability. High accuracy, precision, recall, and F1-score values were observed
on the test sets, indicating that the model effectively distinguishes between
parallelizable and undefined loop structures. The statistical analysis, including
confidence intervals and distribution visualizations, further confirms the
stability and robustness of the learned models.

Beyond overall performance, the study highlighted the importance of considering
application-specific evaluation criteria. While validation error proved effective
for identifying models with strong generalization behavior, the analysis of the
False Positive Rate revealed complementary insights, particularly for scenarios
where incorrect classification of non-parallelizable code as parallelizable may
lead to undesirable consequences. The ability of the proposed approach to achieve
extremely low---and in some cases zero---false positives reinforces its suitability
for practical code analysis contexts.

Despite these promising results, some limitations should be acknowledged. The
scope of this study is restricted to loop-level analysis and does not account for
more advanced loop transformations, such as fusion, unrolling, or tiling. In
addition, although the inclusion of real-world code samples enhances
representativeness, the dataset remains limited to a specific set of programming
patterns and languages.

Future work will focus on extending the analysis to more complex code structures
and transformations, expanding the dataset with additional real-world examples, and
exploring the integration of the proposed classifier into static analysis tools or
development environments. Investigating alternative Transformer architectures and
multilingual code representations also represents a promising direction for further
research.

Overall, the results indicate that Transformer-based models constitute a viable
and effective solution for automatic parallelization potential classification,
offering a robust and scalable foundation for supporting parallel programming
analysis within the defined scope of this study.

\section*{Acknowledgments}

This work was supported by the Coordination for the Improvement of Higher
Education Personnel -- Brazil (CAPES) -- Finance Code 001 through a master's
scholarship. The authors acknowledge the institutional support from the Federal
Rural University of Pernambuco (UFRPE) and the Graduate Program in Applied
Informatics (PPGIA). We also thank the IFROG research group for their valuable
collaboration.

\section*{Author Contributions}

Izavan dos Santos Correia: conceptualization, methodology, software, validation,
formal analysis, investigation, writing original draft, writing review and
editing. Henrique Correia Torres Santos: supervision, review and editing. Tiago
Alessandro Espínola Ferreira: supervision, review and editing.

\section*{Data Availability}

The datasets used and/or analyzed during the current study are available from the
corresponding author upon reasonable request.

\section*{Code Availability}

The code used for experiments in this study is available from the corresponding
author upon reasonable request.

\bibliographystyle{unsrtnat}
\bibliography{references}

\end{document}